\documentclass[apjl]{emulateapj}
\usepackage{apjfonts}

\begin{document}

\title{{\it GALEX} Observations of ``Passive Spirals'' in the Cluster
  Cl~0024+17: Clues to the Formation of S0 Galaxies}
\shorttitle{{\it GALEX} Observations of Passive Spiral Galaxies}
\author{Sean M. Moran$^1$,
  Richard S. Ellis,$^1$, Tommaso Treu$^2$, Samir Salim$^3$, 
  R. Michael Rich$^3$, Graham P. Smith$^{1,4}$, Jean-Paul Kneib$^5$}
\affil{$^1$California Institute of Technology, Department of
  Astronomy, Mail Code 105-24, Pasadena, CA
  91125, USA email: smm@astro.caltech.edu, rse@astro.caltech.edu
  $^2$Department of Physics, University of California, Santa Barbara,
  CA 93106, email: tt@physics.ucsb.edu
  $^3$University of California at Los Angeles, Department of Physics \&
  Astronomy, Los Angeles, CA 90095
  $^4$ School of Physics \& Astronomy, University of Birmingham,
  Edgbaston, Birmingham, B15 2TT, UK. $^5$ Laboratoire d'Astrophysique 
  de Marseille, Traverse du
  Siphon - B.P.8, 13376 Marseille Cedex 12}

\begin{abstract}

We present new results from deep {\it GALEX} UV imaging of the cluster
Cl~0024+17 at $z\sim0.4$. Rest-frame far UV emission is detected from
a large fraction of so-called ``passive spiral galaxies'' --a significant population 
which exhibits spiral morphology with little or no spectroscopic evidence for 
ongoing star formation. This population is thought to represent infalling 
galaxies whose star formation has been somehow truncated by environmental 
processes, possibly in morphological transition to S0 galaxies.  Compared 
to normal cluster spirals, we find that passive spirals are redder in
{\it FUV}--optical 
color, while exhibiting much stronger UV emission than cluster E/S0s - as 
expected for recently-truncated star formation. By modeling the different 
temporal sensitivities of UV and spectroscopic data to recent activity, we show
that star formation in passive spirals decayed on timescales of less than 1~Gyr, 
consistent with `gas starvation' - a process where the cluster environment prevents 
cold gas from accreting onto the spiral disk. Intriguingly, the fraction 
of spirals currently observed in the passive phase is consistent with
the longer period expected for the morphological transformation and 
the subsequent build-up of cluster S0s observed since $z\simeq$0.4. 

\end{abstract}


\section{INTRODUCTION}
In normal spiral galaxies, the ultraviolet (UV) luminosity depends 
sensitively on the presence of two components: young, hot stars and
dust. UV continuum observations are sensitive to star formation 
on timescales of $<$ few 100~Myr and can thus place strong constraints 
on both current and recent activity, particularly if independent estimates of 
dust extinction are available. Due to its wide field of view, the Galaxy
Evolution Explorer ({\it GALEX}) satellite allows an efficient census of the 
UV content of galaxies \citep{martin05}. 
Here we use {\it GALEX} imaging to address a key issue
in galaxy evolution: the search for objects in morphological transition
in intermediate redshift clusters.

It has been known for many years that galaxies in rich clusters are generally 
bluer and more active at intermediate redshift \citep{bo78, bo84}.
With the aid of Hubble Space Telescope ({\it HST}) imaging, recent quantitative
studies of the evolving morphology-density relation \citep{d97, smith05,
postman05} lend support to the suggestion that blue star-forming cluster spirals have 
been somehow transformed into red S0s \citep{couch87}. In order to understand how such 
transformations might occur, several spectroscopic and photometric surveys of
intermediate redshift clusters have sought to pinpoint samples of galaxies 
which are in the (presumed short-lived) {\it transitional phase} 
\citep[e.g.][]{couch98,pogg99,balogh99}.

Several candidate object classes have been identified from
spectroscopic data utilizing the combination of [O II] emission and Balmer
absorption, which are respectively sensitive to ongoing star formation
and recent star formation on timescales of $\simeq$1 Gyr. These
include the ``E+A'' or ``k+a'' galaxies that exhibit a deep
Balmer H$\delta$ line \citep{dressler83, dressler92}, and the ``e(a)''
galaxies that combine [O II] emission with strong Balmer
absorption \citep{pogg99}.  Recent studies indicate that e(a) spectra
are largely associated with dusty starburst galaxies
\citep{poggianti00}, and that (k+a)s are associated with the 
post-starburst remnants of a merger or close encounter 
(Goto 2005; Tran et al. 2003--but see also Poggianti et al. 2004). 
However, there are few intermediate redshift S0 galaxies with
e(a) or post-starburst spectral signatures \citep{pogg99}. Indeed
\citet{moran05} found no difference between the stellar populations of
Es and S0s at $z=0.4$- an unexpected result if S0s were recently created 
from starbursting systems. Clearly, our understanding of the
evolutionary link between e(a)/k+a galaxies and S0s remains
incomplete.

Recently, a further interesting class has been identified -
the so-called ``passive spirals'' \citep{couch98, d99, pogg99,
goto03}.  These objects exhibit spiral morphology in {\it HST} images, but
reveal weak or no [O II] emission. 
Some authors have suggested that these are spiral-to-S0
transition objects where cessation of star formation occurs on a
faster timescale than the transformation of spiral morphology. Such 
a delay in the morphological transformation would naturally explain 
the older stellar populations inferred from optical spectra of 
S0s \citep{pogg99, moran05}. Theoretical models by \citet{bekki02}
have shown that this scenario is consistent with gas ``starvation'', 
an interaction with the intra-cluster medium that serves to inhibit 
star formation by halting the accretion of cold halo gas onto the 
galaxy disk \citep[see][]{larson80, quilis00, tt03}.  Without further 
gas accretion, Bekki et al. (2002) found that the spiral arms fade within
$\simeq$ 3 Gyr.

\citet{goto03} find that passive spirals in the SDSS reside preferentially in
intermediate density environments, confirming this is a
cluster-related phenomenon. Balmer H$\delta$ absorption is weaker in
passive spirals than in the overall spiral population \citep{goto03,
pogg99}, suggesting their stellar populations already resemble the
older populations found in E+S0 (``early-type'') galaxies. To date, however, 
little else is known about the detailed star formation histories of passive 
spirals.

In this paper, we combine {\it GALEX} UV imaging of Cl~0024+17, a rich
cluster at $z=0.4$, with extensive {\it HST} imaging and Keck
spectroscopy to constrain the recent and ongoing star formation rates (SFRs)
of passive spirals. We compare our measurements of UV--optical colors 
and key spectral lines with model star formation histories. This 
combination of diagnostics allows us for the first time to distinguish 
between several explanations for the nature of the passive spirals.

\section{OBSERVATIONS AND SAMPLE SELECTION}

\subsection{Data}

\begin{figure}
\centering
\includegraphics[width=\columnwidth]{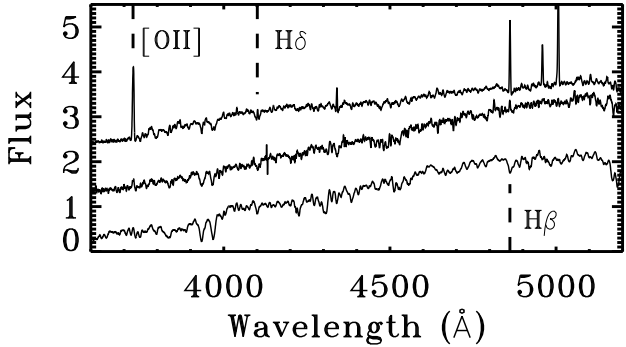}
\caption{\label{fig:coadded} Normalized coadded spectra of E+S0
  galaxies (lower spectrum), passive (middle) and active (upper) spiral
  galaxies, offset vertically, in arbitrary units, 
  created as in Moran et al. (2005). Key spectral lines are marked.
}
\end{figure}

We make use of {\it HST} imaging and Keck spectroscopy of Cl~0024+17
from the comprehensive wide-field survey fully described in
\citet{tt03} and \citet{moran05}. Briefly, the survey builds on a 
sparsely-sampled
mosaic of 39 WFPC2 images taken in the {\it F814W} filter ($\sim I$ band),
providing coverage to a projected radius $> 5$ Mpc. \citet{tt03}
reported reliable morphological classifications to $I=21.1$ and we
combine these classifications with Keck spectroscopy in order to
define three samples of cluster members: passive spiral, ``active'' or
star-forming spiral, and E+S0.

Cl~0024+17 was observed for 13.3ksec with {\it GALEX} in 2004 October 
(GO-22; Cycle 1; PI Treu) in both near ({\it NUV}) and far ultraviolet
({\it FUV}) filters \citep{martin05, morrissey05}.  As {\it GALEX}'s 
field of view is $\sim1\fdg2$, the images readily cover the full {\it HST}
mosaic. The {\it NUV} band closely matches the rest-frame {\it FUV}
at $z\simeq$0.4 ($\lambda_c\sim1620\mbox{\AA}$), so we focus on that
band. Galaxy fluxes were measured within $6\arcsec$
circular apertures, centered on the optical position, and comparable
to our measured {\it NUV} FWHM of $\simeq5\farcs5$. Such a fixed-size
aperture makes the detection limits easier to interpret. We apply 
a fixed 20\% aperture correction for agreement with
SExtractor-derived total magnitudes ({\tt MAG\_AUTO}), and for comparison to
{\tt MAG\_AUTO} magnitudes in {\it F814W} \citep{bertin96}.
The {\sl kcorrect} software v.4\_1\_2 \citep{blanton03} was used to convert
observed {\it NUV} and {\it F814W} fluxes to rest-frame {\it FUV} and
{\it V} luminosities. A Galactic extinction of 
E({\it B--V})=0.056 \citep{schlegel98} was assumed.

Observations with the DEIMOS spectrograph on Keck II from October 2001
to October 2005 secured spectra for over $500$ members of Cl~0024+17
(300 with {\it HST} imaging).
Details are provided in \citet{tt03}, \citet{moran05}, and Moran et
al. (2006, in prep). We augment our sample with  cluster members 
derived from the CFHT survey of \citet{czoske01}, who
reported the strengths of (or the absence of) several spectral
lines including [OII].  For both the Keck and CFHT spectra, we use
Lick-style indices which are measured as described in \citet{moran05}
and references therein.

\subsection{Sample Selection}

As faint emission line objects are easier to robustly identify 
than those without emission, there is a 
potential bias against identifying ``passive'' objects. To
minimize this effect, we adopt a bright magnitude limit $F814W
\le 21.1$, where the sample is nearly unbiased
\citep{moran05}. Reliable visual morphologies are available to
the same limit\citep{tt03}. 

We define as ``passive'' any Sa--Sd spirals which have EW([OII]) $> -5
\mbox{\AA}$\citep{d99, pogg99}, with ``active'' spirals having 
EW([OII]) $\le -5\mbox{\AA}$; a more stringent limit is precluded by 
signal to noise considerations. While this definition does
not exclude (k+a) or e(a) spirals as separate classes, our 
Cl~0024 sample contains only two spirals with 
EW(H$\delta$)$\equiv $H$\delta_A > 5 \mbox{\AA}$, both with [OII] emission.
To be consistent in comparing passive spirals with
E+S0 galaxies, we likewise exclude all early-types with EW([OII])$
< -5\mbox{\AA}$  ($\sim25\%$).  
We manually examined all optical and UV galaxy images
to ensure correct associations. We removed from the sample all
galaxies where the UV flux is centered on a neighbor, or is
otherwise contaminated; this reduced the total sample by
$12\%$. Adopting these procedures, we obtained a sample of 68 cluster
spirals, of which 24 are passive and 44
active. The comparison sample of passive E+S0s contains 75
objects. In Figure~\ref{fig:coadded}, we show the coadded, normalized
spectra of E+S0s, passive spirals, and active
spirals.

\section{UV EMISSION IN PASSIVE CLUSTER SPIRALS}

\begin{figure}
\centering
\includegraphics[width=\columnwidth]{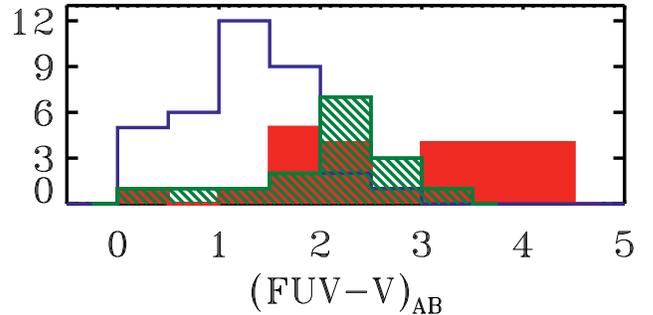}
\caption{\label{fig:nuvcolors} 
   The distribution of rest-frame
  ({\it FUV--V})$_{\mathrm{AB}}$ colors for active spiral (blue), passive spiral
  (hatched green), and early-type (solid red) galaxies. 
}
\end{figure}

While the optical spectra of passive spirals are noteworthy for their
resemblance to those of red early-types \citep[Figure~1,][]{d99},
their {\it FUV} properties 
reveal a very different picture. A large fraction of passive spirals
in Cl~0024 emit vigorously in the {\it FUV}. The fraction of UV-detected
passive spirals ($>3\sigma$) is $67\pm 17\%$, not significantly
different from that of active spirals ($80\pm13\%$). 
However, only $25\pm7\%$ of the early-type galaxies are similarly 
detected in the UV, clearly indicating that the stellar populations of 
passive spirals differ from those of the early type galaxies. 

Moreover, Figure~\ref{fig:nuvcolors} shows that the UV-detected 
passive spirals have {\it FUV~--~V} colors 
intermediate between the bluer active spirals and the early-types, 
which span a wide range extending the furthest to the red. A K-S test
confirms that the {\it FUV~--~V} colors of passive spirals differ from those
of E+S0s with $>99\%$ confidence. These intermediate colors 
strongly suggest that passive spirals bridge the gap between the 
blue and red galaxy sequences. Below, we evaluate 
this suggestion by considering model star formation histories. 

\section{MODEL STAR FORMATION HISTORIES}

\begin{figure}
\centering
\includegraphics[width=\columnwidth]{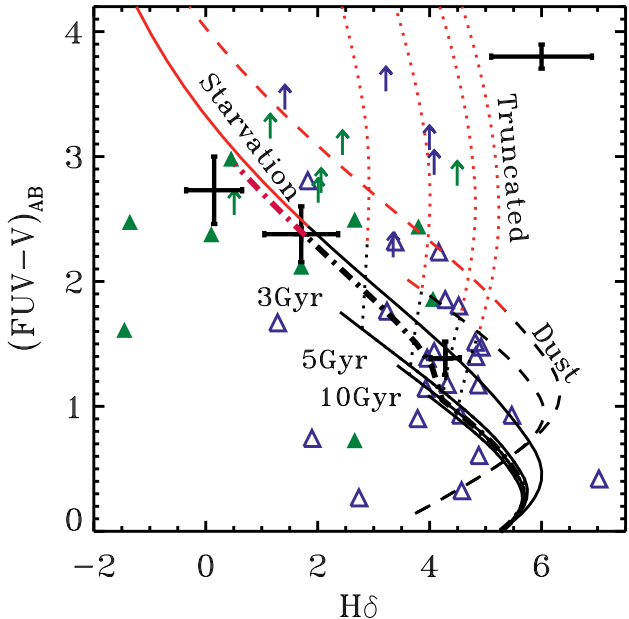}
\caption{\label{fig:hd_uv} ({\it FUV~--~V})$_{\mathrm{AB}}$ color, rest frame, 
  versus H$\delta_A$, for passive (filled green triangles) and active
  (open blue triangles) spirals, with typical error at
  upper right. Arrows indicate lower limits on
  {\it FUV~--~V}, color coded in the same manner. 
  Black crosses indicate, from lower right to upper left, the mean 
  location of the active spirals 
  (with error on the mean), the same for passive spirals, and a composite of
  13 passive spirals that cannot be plotted individually.
  Solid, dotted, dashed, and dash-dotted lines respectively mark several
  classes of model star formation histories, as labeled on the plot
  and discussed in the text. Tracks are colored
  red for regimes where EW([OII])$> -5\mbox{\AA}$. Model ages increase 
  generally from the lower right to the upper left.}
\end{figure}

The detection of strong {\it FUV} emission from passive spirals is perhaps
surprising in a simple picture where the UV light and optical emission
lines are generated simultaneously from the same star forming
regions. Here, we explore possible explanations.
The key to understanding the observation lies in the
hierarchy of lifetimes of the stars responsible for the strong
H$\delta$ absorption (10$^9$ yrs), {\it FUV} flux (10$^{7}$-10$^{8}$
yrs) and hard ionizing photons (10$^{7}$~yrs).

It is useful at this point to introduce the {\it FUV~--~V} vs H$\delta_A$
diagram, which we will use as our main diagnostic tool. Firstly, 
by comparing signatures of star formation over different
timescales, this diagram allows us to explore the recent star
formation history of active and passive spirals and to assess whether
they can be connected in an evolutionary sequence. Secondly
-- since {\it FUV~--~V} is sensitive to dust extinction while H$\delta$ is not
\citep[e.g.][]{macarthur05, sato06} -- this
diagram allows us to explore whether the redder {\it FUV~--~V} colors of
passive spirals with respect to active spirals could be due to dust
instead of differences in the recent star formation history.

Figure~\ref{fig:hd_uv} shows the location of passive and active
spirals in the {\it FUV~--~V} vs H$\delta_A$ diagram. Active spirals
(open blue triangles) are clustered at a blue {\it FUV~--~V} color 
with moderately strong Balmer absorption (H$\delta\sim 4\mbox{\AA}$), 
whereas the passive spirals (solid green triangles) extend to 
redder {\it FUV~--~V} colors with weaker H$\delta$. The mean location 
of each population is marked as an error-barred point; errors indicate 
uncertainty in the mean. The error-barred point with the weakest H$\delta$ 
is a composite of 13 passive spirals that cannot be plotted 
individually, either due to lack of reliable H$\delta$ or
contamination of FUV flux. We coadded available spectra for these
galaxies to measure H$\delta$, and we adopt the mean of available {\it
  FUV~--~V}. This composite lies near the locus of individually
plotted passive spirals at low H$\delta$ and red {\it FUV~--~V}. 

Evolutionary tracks from the population synthesis code of \citet{bc03}
are overlayed in Figure~\ref{fig:hd_uv} We consider only solar
metallicity, and only plot model points reached in less than the time
between formation at $z\sim5$ and observation at $z=0.4$ ($\sim8$~Gyrs). 
Tracks are in red for regimes where EW([OII])$ > -5 \mbox{\AA}$;
we use SFRs and $L_B$ from the \citet{bc03} models to estimate EW([OII])
according to the relation in \citet{barbaro97}. We calculate
[OII] attenuation by dust following \citet{calzetti94}, with
nebular emission extincted more strongly than the stellar
continuum ($E_*/E_{g}\sim0.5$), but caution that
this is only a representative model. We subtract $0.9\mbox{\AA}$ from 
H$\delta$ for all model tracks to account for infilling by nebular 
emission, as estimated from \citet{barbaro97}.

Solid lines represent
exponentially-declining SFRs. Those marked with characteristic timescales 
$\tau=3,5,10$~Gyr have been found to reproduce the optical spectra 
of Sa, Sb, and Sc--type galaxies (``active'' spirals), respectively 
\citep{poggianti96}.  Dashed lines include dust extinction of $A_V=0.6$, 
corresponding to the mean difference in {\it FUV~--~V} between 
the passive and active spirals, for a Cardelli et al. (1989)
extinction law with R$_V$=3.1. The lower dashed line is a representative
$\tau=5$~Gyr model with dust. The uppermost dashed and solid lines
represent a $\tau=1$~Gyr SFR (with and without dust) and will be 
discussed below. Active spirals largely occupy the 
region in between tracks with zero and moderate extinction.

{\it While adding extinction to the active spiral models can 
serve to reproduce the lack of emission lines and the 
redder FUV~--~V colors of the passive spirals, as expected it has very
little effect on the H$\delta$ values}.  Thus, while a model with
extincted star formation is sufficient to explain those passive
spirals with higher H$\delta$, it is unable to account
for the bulk of the population observed with weaker
H$\delta$. Conceivably, more complex dust screens could
account for this deficiency, but we will 
explore a more straightforward explanation where the SFR is 
suppressed as a result of environmental processes. 

We first consider ``truncated'' star formation:  in
Figure~\ref{fig:hd_uv},  we plot (as dotted lines)
several representative $\tau=3-10$ Gyr models where SFR instantaneously falls 
to zero at an age between $t=5-7$~Gyr, representing a range of
initial formation ages. Such models eventually generate both lower
H$\delta$ values and redder {\it FUV~--~V} colors after star formation
is turned off, but the UV emission decays
too rapidly, falling below our {\it GALEX} detection limit well before
low H$\delta$ values are reached. 

Only models with a relatively fast 
exponential decline in SFR ($\tau\le1$~Gyr, upper solid and dashed lines in 
Figure~\ref{fig:hd_uv}) can match the {\it FUV~--~V} and H$\delta$ of 
the bulk of passive spirals. Alternatively, any of the active spiral 
models with 
$\tau=3-10$~Gyr can migrate toward the $\tau=1$~Gyr track if its 
star formation begins to decline at a much steeper exponential rate 
(thick dash-dotted line in Figure~\ref{fig:hd_uv}), entering into a
spectroscopically passive phase. The two scenarios are essentially equivalent,
differing only in the age of the galaxy when it enters the passive
phase, and the amount of dust extinction applied.

\section{Discussion}

Remarkably, the star formation histories with a sharp decline in SFR 
resemble those expected for spirals affected by ``starvation''
\citep{bekki02,larson80, quilis00}.  As the accretion of gas is
rapidly cut off, star formation decays on a short timescale as the
remaining disk gas is used up ($0.8 $~Gyr$ < \tau < 2 $~Gyr, according to
Larson et al. 1980). More generally, we might expect that
any ``gentle'' physical mechanism that inhibits star formation without
immediately affecting the spiral morphology would behave similarly.
A gentle mechanism is also easier to
reconcile with the observed homogeneity of the stellar population and
dynamical properties of S0 galaxies \citep{tt03,moran05}. Models with 
no explicit break in SFR, but with low values of $\tau$ ($\sim1~Gyr$),
reproduce the passive spiral data as well as the starvation model, and 
seem not to require the action of any physical mechanism. Yet passive 
spirals seem to be most prevalent in clusters \citep{goto03}, suggesting that a
cluster-related mechanism contributes to their formation.

Since few passive spirals are observed with an {\it FUV} upper limit
(8/24), we might expect that any morphological transition occurs soon after a
passive spiral fades below our {\it FUV} limiting magnitude.  For the model
starvation track plotted as a thick dash-dotted line in 
Figure~\ref{fig:hd_uv}, the total
lifetime from the halt of gas accretion to the point where the {\it FUV}
luminosity fades below our detection limit is $\sim3$~Gyrs, in
good agreement with the morphological transformation
timescale predicted by \citet{bekki02}.

This particular evolutionary track spends $\sim1.5$~Gyr in a phase
with EW([O II])$ \ge -5.0\mbox{\AA}$ and detectable {\it FUV}
(for the mean $V_{AB}=20.6$). The fraction
of all bright spirals in this UV-detected, spectroscopically passive
phase is $27\pm7\%$ in Cl~0024, though some contamination is likely
from galaxies undergoing dust-obscured star formation. These
statistics enable us to consider whether the fraction of passive
spirals observed is consistent with the buildup of the S0
fraction from $z=0.4$ to $z=0.0$. 

Following \citet{smith05}, we define the ratio of S0 to ellipticals
today, $N_{S0, z=0}/N_{E, z=0}$, in terms of the passive spiral
fraction and S0 fraction at $z=0.4$:
\begin{equation}
  \frac{N_{S0, z=0}}{N_{E, z=0}}=\left( \frac{N_{S0, z=0.4}}{N_{E,
  z=0.4}} + \frac{\Delta t}{\tau} \frac{N_{psp, z=0.4}}{N_{E,
  z=0.4}} \right) \frac{N_{E, z=0.4}}{N_{E, z=0.4}+\Delta N_E}
\end{equation}
where $\Delta t$ is the lookback time to $z=0.4$, 4~Gyr, and $\tau$ is
the passive spiral lifetime of 1.5~Gyr. The last term accounts for the
buildup of ellipticals through other means, which we set equal
to 1 for this simple analysis. Taking $N_{S0, z=0.4}/N_{E,z=0.4}=0.65$
\citep{d97} and $N_{psp, z=0.4}/N_{E,z=0.4}=0.25$\footnote{According
to \citet{d97}, the ratio of ellipticals to spirals is nearly unity in
Cl~0024, so $N_{psp, z=0.4}/N_{E,z=0.4} \simeq N_{psp}/N_{sp}$}, we
calculate $N_{S0, z=0}/N_{E, z=0}=1.3$, in rough agreement with
$N_{S0, z=0}/N_{E, z=0}\sim1.7\pm0.6$ from \citet{d97}. This
shows that passive spirals undergoing starvation can evolve into 
present day S0s, consistently with the observed
evolution of the morphology density relation. 

While starved spirals cannot build up the {\it entire} population of local
S0s due to differences in the stellar mass functions of local S0s and
spirals \citep{kodama01}, our result 
provides strong evidence that the abundant passive spiral population at 
intermediate redshift is an important transition population which 
contributes to the present-day cluster S0 population. By combining 
the different post-burst timescales probed by {\it GALEX} UV imaging 
and optical spectroscopy, we have delineated a duty-cycle of gradual 
decline in activity consistent with the recent growth in the S0 population.

\begin{acknowledgements}
We thank L. MacArthur and I. Smail for valuable comments.
RSE acknowledges financial support
from NSF grant 
AST-0307859 and STScI grants HST-GO-08559.01-A and HST-GO-09836.01-A. Financial support from NASA for {\it GALEX} GO-Cycle 1 program 22 is 
acknowledged. 
\end{acknowledgements}

\end{document}